\documentstyle[aps,prb,twocolumn,epsf,rotate]{revtex}

\begin{document}

\twocolumn[\hsize\textwidth\columnwidth\hsize\csname
@twocolumnfalse\endcsname

\title{Phonon-mediated drag at $\nu=1/2$: A test of the Chern-Simons
composite fermion theory}

\author{Martin C. B{\o}nsager$^{1}$, Yong Baek Kim$^{2}$, and A. H. MacDonald$^{1}$}
\address{$\mbox{}^1$ Department of Physics, Indiana University,
Bloomington, Indiana 47405}
\address{$\mbox{}^2$ Department of Physics, The Ohio State University, Columbus, 
Ohio 43210}
\date{\today}
\maketitle

\begin{abstract}

We report on a study of the phonon-mediated frictional drag between
two-dimensional electron layers at the Landau level filling factor $\nu=1/2$ 
predicted by Chern-Simons composite-fermion theory.
Frictional drag between widely spaced layers is dominated by the phonon-mediated interaction
and is altered from its zero-field form
because of the large composite-fermion effective mass and, less strongly,
by magnetic local-field effects.  We find that 
the reduced drag resistance $\rho(T)/T^2$ peaks at a 
larger temperature and has a different density dependence than 
at zero field. We compare these results with recent experimental findings.

\end{abstract}


\vskip2pc]


\section{Introduction}

Comparisons between measured transport properties and theoretical predictions
have been fruitful in providing important information about elementary 
excitations in electronic systems. 
A recent and rather unusual example may be
found in studies of the frictional drag  between
separately contacted, nearby, two-dimensional electron layers.  
Because it is a measure of the interlayer scattering rate, drag
resistance is sensitive to Coulomb and phonon-mediated interactions 
between the layers and is related to correlated density fluctuations 
in the coupled bilayer system.  Since these depend crucially on 
underlying electronic degrees of freedom, frictional drag 
can serve as an important test of specific theoretical
predictions concerning the fluctuation spectrum of a given electronic system.    
At zero field, these studies have verified many qualitative features
predicted by the random phase approximation (RPA) theory of
correlations in moderate density two-dimensional electron systems.
On the other hand, they have also demonstrated 
the importance, at a quantitative level, of various corrections.
These studies have a greater potential importance
in the fractional quantum Hall regime, where simple perturbative RPA-like theories
fail and our understanding of correlations is much less complete.
In the fractional quantum Hall regime, all electrons share the same
quantized kinetic-energy and low-energy properties are entirely determined
by electron-electron interactions. Electrons in the fractional
quantum Hall regime are in the limit of extreme correlations. 
It is the failure of perturbation theory
which admits all the peculiarities and surprises which have been discovered 
in this field over the past two decades.
  
One example of a peculiar strong-correlation consequence is 
presented by the Fermi-liquid-like properties of the compressible states which occur when
the lowest Landau level is half-filled.  Experimental evidence then supports 
the existence of gapless fermionic excitations which form
a Fermi surface.\cite{stormer}
This is particularly surprising because the kinetic energy of the electrons 
in the lowest Landau level is quenched by the large magnetic fields, so
this Fermi surface must result purely from interaction effects.
On the theoretical front,
the Chern-Simons composite fermion theory has been spectacularly
successful in explaining various experimental findings at the 
phenomenological level \cite{hlr,jain}.
It starts from a singular gauge transformation
which attaches two flux quanta to each electron.
These new `composite particles', like electrons, obey Fermi statistics 
and are therefore called composite fermions.
At $\nu=1/2$, the mean-field flux-density from the attached flux quanta exactly 
cancels that coming from the external magnetic field.
Neglecting fluctuations, composite fermions at $\nu=1/2$ therefore
experience zero magnetic field and have a Fermi surface.
In the random phase approximation (RPA),
the density-density correlation function for composite fermions
is equal to the density-density correlation function for completely 
spin-polarized electrons at zero magnetic field.
However, the density-density correlation function of the physical 
electrons contains an additional local field term associated with the 
attached flux quanta.  
We will refer to this additional term as the magnetic local field.
In RPA theory, the magnetic-field dependence of the correlation
function is completely embedded in this quantity.

Chern-Simons theory has the advantage of being technically and
conceptually simple.  It suffers, however, from various problems, most
associated with the composite fermion mass scale $m^*$.
On physical grounds, it is known that $m^*$ is determined by the 
interaction energy scale in the lowest Landau level.
However, the theory does not have a proper lowest Landau level projection
so that $m^*$ has to be fixed phenomenologically. 
Unfortunately, this naive approach leads to a violation of the f-sum rule and 
Kohn's theorem\cite{kohn61} in the sense that certain response
functions should contain the bare electron mass $m_{b}$. 
A scheme to correct for this deficiency was developed by
Simon and Halperin\cite{simon93} who introduced the
Modified Random Phase Approximation (MRPA). The MRPA includes
a Fermi Liquid like correction, and allows for an effective
mass $m^*$ different from the band mass $m_{b}$ while still satisfying the
f-sum rule and Kohn's Theorem.
Even though this prescription turns out to be surprisingly successful,
a more microscopic understanding of the role of the lowest Landau level
projection in the theory is desirable.
  
Recently several attempts have been made to construct a truly
lowest Landau level theory of composite 
fermions.\cite{read94,shankar97,lee98,pasquier98,stern98,read98,shankar99}
Inspired by $\nu=1/2$ variational wavefunction\cite{read94}, it has been realized that the 
position of an electron and a nearby zero of the wavefunction (called 
a `vortex') are displaced
by $k_i l_B^2$ for each electron labeled by $i = 1, \cdots, N$.
The lowest Landau level constraint and strong correlations tend
to bind vortices to electrons.
The composite object that consists of an electron and two vortices 
satisfies fermionic statistics and, as a result, the $k_i$'s
have to be chosen differently for each composite object.
The ground state of these `neutral composite fermions'
is given by a Fermi sea in the space of $k_i$.
One can also show that $k_i$ acts as the guiding-center-translation generator 
in the lowest Landau level.  Within a Hartree-Fock theory of these 
composite objects, the effective kinetic energy 
has the form $k^2_i/2m^*$ with an effective mass $m^*$ 
given correctly by the interaction energy 
scale \cite{read94,shankar97,lee98,pasquier98,stern98,read98,shankar99}. 
A self-consistent theory of neutral dipolar composite fermions has been constructed
and it has been shown that, in the long-wavelength low-energy limit,
this theory (taking the 
bare electron mass to infinity to reflect the absence of a kinetic energy 
term in the lowest Landau level Hamiltonian) gives the same
physical response functions as those of the Chern-Simons composite
fermion theory \cite{stern98,read98}.
Coulomb mediated frictional drag between two $\nu=1/2$ systems depends mostly on
the long wavelength and low energy limits of density-density correlation
in each layer. As a result, the two theories mentioned above should make 
identical predictions for Coulomb drag.  This quantity has been 
measured \cite{lilly98} and the observations are consistent at relatively
high temperatures with the theoretically 
predicted $T^{4/3}$ power law.\cite{ussishkin97,sakhi97,kim96}
The data does, however, exhibit a low temperature anomaly that 
has not yet been fully understood.\cite{ussishkin98,zhou99}

When two 2D electron systems are widely separated, the contribution 
to drag from Coulomb scattering between the layers is suppressed and
phonon mediated interaction dominates.
Unlike the Coulomb drag case, 
phonon mediated drag is sensitive to density-density correlations 
in each layer at relatively high frequencies and at wavevectors comparable
to the inverse interparticle separation. 
Therefore studies of phonon mediated drag should provide useful 
information about excitations at finite frequencies and at
large wavevectors, leading to a test of the 
underlying theory at many different energy and length scales,
not just at the long wavelength and low energy limits. 
Since the equivalence between Chern-Simons theory and 
the neutral dipolar composite fermion theories has been 
established only in the long wavelength and low energy limits,
studies of phonon drag may reveal important differences between 
these two theories.
As the first step to investigate the validity of these theories,
in this paper we report on a comparison between recent measurements of 
frictional drag\cite{zelakiewicz} between widely separated two-dimensional 
electron layers at $\nu =1/2$ and Chern-Simons composite fermion theory 
predictions. We find some qualitative differences between observations and 
trends predicted by theory and speculate on their implications for the 
theory of the half-filled Landau level.

In the following section we review the theory of frictional drag and discuss 
some of the features that are important in analyzing drag at $\nu=1/2$.
The theoretical framework we use to describe frictional drag is based on
earlier $B=0$ work involving two of us.\cite{bonsager98}  While there are some
issues in the theoretical literature\cite{tso92,zhang93,badalyan99}
on frictional drag which are not completely settled, these do not
impact on the conclusions we draw here and we do not discuss them
further.  In Sec. \ref{phonondrag} we present numerical results for drag, 
calculated from (M)RPA composite-fermion theory, and compare these with recent
experimental results obtained by Zelakiewicz {\it et al.}\cite{zelakiewicz}.
Our conclusions are presented in Sec. \ref{conclusions}.

\section{Frictional drag}

When two or more electronic systems are placed in close proximity, their
transport properties are interdependent.
In particular, currents in one subsystem can induce (or 'drag') currents in other
subsystems.  The first drag experiments\cite{solomon89} were on coupled
two-dimensional and three-dimensional electron system.  Drag between two-dimensional
layers has been measured in
electron-electron\cite{gramila91,gramila93}, electron-hole\cite{sivan92}, and
recently in hole-hole\cite{jorger} systems.
In this paper we consider frictional drag between two two-dimensional
electron gases embedded in a GaAs-AlGaAs double quantum well
system.  The quantum wells widths $L$  are $\sim 200$ \AA\ and are
separated by a center-to-center distance $d$ which is large enough to ensure that
interlayer tunneling can be neglected ($d\gtrsim 300$ \AA).
We imagine a current density $J_1$ being drawn in the first layer while the second
layer is an open circuit.  To counter balance the stochastic drag force
between the layers, an electric field $E_2$ will build up
in the second layer.  The transresistivity, $\rho_{21}$, is defined as
\begin{equation}
\rho_{21} \equiv E_2/J_1,
\end{equation}
$\rho_{21}$ depends on the strength of interlayer interactions and on the
phase space available for interlayer scattering events.

Theoretical expressions for $\rho_{21}$ can be derived by conventional
linear response theory at various levels of
sophistication.\cite{jauho93,zheng93,kamenev95,flensberg95}
Under conditions which are met in current experiments,
the transresistivity is given by
\begin{eqnarray}\label{rho21}
\rho_{21}&=&\frac{-\hbar^2}{8\pi^2 e^2 n_1n_2k_{\rm B}T}\int_0^{\infty}dq\ q^3
\nonumber\\
&&
\int_0^{\infty}d\omega \left| \frac{W_{21}(q,\omega)}{\varepsilon(q,\omega)}\right|^2
\frac{{\rm Im}\Pi_1(q,\omega){\rm Im}\Pi_2(q,\omega)}{\sinh^2(\hbar\omega/2k_{\rm B}T)}.
\end{eqnarray}
Here $n_i$ and $\Pi_i(q,\omega)$ are the two-dimensional electron density
and the polarization function, respectively, of layer $i$.  In
Eq. (\ref{rho21}) $W_{ij}(q,\omega)$ is
the reciprocal space interaction between layers $i$ and $j$,
and $\varepsilon(q,\omega)$ is the interlayer screening function, given
by
\begin{equation}\label{epsilon}
\varepsilon=[1+W_{11}\Pi_1][1+W_{22}\Pi_2]-W_{21}^2\Pi_1\Pi_2.
\end{equation}
Given the interlayer interaction, this theory of frictional drag depends only
on the polarization functions of the 2D electron layers.

Both interlayer Coulomb interactions and phonon-mediated interactions
contribute to interlayer friction:
\begin{equation}\label{contributionsum}
W_{ij}(q,\omega)=U_{ij}(q)+{\cal D}_{ij}(q,\omega).
\end{equation}
Taking account of the finite quantum well widths, the Coulomb contribution is 
\begin{equation}
U_{ij}(q) = \frac{e^2}{2\kappa q}B_{ij}(qd,qL/2)
\end{equation}
where $\kappa$ is the dielectric constant of the semiconductor and
$B_{ij}$ is a width-dependent form factor\cite{bonsager98}. 
The phonon mediated interaction is given by
\begin{equation}
{\cal D}_{ij}(q,\omega)=\sum_{\lambda}\int\frac{dQ_z}{2\pi\hbar}
|M_{\lambda}({\bf Q})|^2 F_i(Q_z)F_j(-Q_z)D_{\lambda}({\bf Q},\omega)
\end{equation}
where $F_i(Q_z)=\int dz |\varphi_i(z)|^2 e^{-iQ_Zz}$ is the Fourier
transform of the electron density in the direction perpendicular to the
layers ($\varphi_i$ is the subband wavefunction in layer $i$),
and $D_{\lambda}({\bf Q},\omega)$ is the phonon propagator.
We use the notation ${\bf Q}=({\bf q},Q_z)$.
The index $\lambda$ denotes longitudinal and transverse phonon modes,
$\lambda=l,t$.
Both acoustic and optical phonons can contribute to drag, but in
GaAs/AlGaAs systems acoustic phonons dominate under the conditions
studied in this paper.\cite{hu98,guven97}
The electron-acoustic phonon coupling matrix elements are given
by\cite{price81,lyo88}
\begin{equation}
|M_l({\bf Q})|^2=\frac{\hbar Q}{2\varrho c_l}\left[ D^2 +
(eh_{14})^2\frac{9q^4Q_z^2}{2Q^8}\right]
\end{equation}
\begin{equation}
|M_t({\bf Q})|^2=\frac{\hbar}{2\varrho c_t}(eh_{14})^2
\frac{8q^2Q_z^4+q^6}{4Q^7}
\end{equation}
where the $c_{\lambda}$ are the phonon velocities for longitudinal and transverse
phonons, $\varrho$ is the mass density of GaAs/AlGaAs, $D$ is the deformation
potential, and $eh_{14}$ is the piezoelectric constant.  In our numerical
evaluations we use $c_l=5140$ m/s, $c_t=3040$ m/s, $\varrho=5300$ kg/m$^3$,
$D=-13.0$ eV, and $eh_{14}=1.2\times 10^9$ eV/m.

Because their contributions come dominantly from different regions of
wavevector and frequency, it is possible,  both experimentally and
theoretically, to separate Coulomb and phonon-mediated contributions to drag.
Coulomb drag comes from $q\lesssim d^{-1}$,
whereas phonon-mediated drag is dominated by contributions from $q\simeq2k_F$
and frequencies at the peak in ${\cal D}_{ij}$ near $\omega=c_{\lambda}q$.
Work at zero magnetic field has established that
Coulomb drag dominates at small interlayer separations($d\sim 300$ \AA)
but falls off rapidly with increasing $d$.  Phonon mediated drag, on
the other hand, depends only weakly on $d$ and has been observed for
layers separated by distances approaching $1 \mu$m.

\subsection{Frictional drag at $B=0$}

Frictional drag in the absence of a field is now well
understood and the extensive literature has recently been
reviewed.\cite{rojo99}  Eq. (\ref{rho21}), combined with the
Random Phase Approximation for the 2D electron system
polarization function, successfully\cite{gramila91} predicts a low-temperature
drag $\rho_{21}\propto T^2$, a substantial
enhancement\cite{flensberg94,hill97,noh98}
of $\rho_{21}$ due to the excitation of plasmon modes at
around half the Fermi temperature, and a crossover from
Coulomb drag to phonon-mediated drag for layers separated by
more than $\sim 50 {\rm nm}$.  Phonon-mediated drag is readily identified
by its signature dependences on the ratio of electron densities in the
two layers and on temperature.
For phonon-mediated drag $\rho_{21}\propto T^6$ at low temperatures ($T\sim 0.1$ K)
and varies approximately linearly at higher temperatures ($T\gtrsim 5$ K).
As a consequence the scaled
transresistivity $\rho_{21}/T^2$, which is more nearly constant in the
Coulomb case, will have a well defined peak.  The origin
of this peak in $\rho_{21}/T^2$ is easily understood by looking at
Fig. \ref{phasespace} which shows the phonon resonance and the particle-hole
continuum in phase space.  The integrand in Eq. (\ref{rho21}) is
cut off exponentially with $\omega$ at $\hbar\omega\sim k_{\rm B}T$.
The full phonon resonance is thus exploited when
$T=T_{\rm peak}\sim \hbar c_{\lambda}2k_F/k_{\rm B}$, and a further increase
in the temperature will only lead to a relatively smaller increase
in $\rho_{21}$.

The occurrence of a
peak depends only on the phonon resonance and the existence of a
Fermi surface, which leads to a sharp edge of the particle-hole continuum
as shown in Fig. \ref{phasespace}.  The precise location of the peak, however,
will depend on detailed properties of the polarization function at large $q$ as
well as on many other parameters of the experiment.
At $B=0$, good agreement with experiment for the overall shape of 
scaled transresistivity curves and
for the location of their peaks corroborates the theory.
Disagreements in magnitude can in part be attributed to
uncertainties in the electron-phonon interaction parameters ($\rho_{21}$
varies approximately as the fourth power of the deformation potential
constant $D$ and this parameter is not accurately known).  
Corrections to the RPA polarization function
at large wavevector\cite{bonsager98} may also play a role.
Moreover, the magnitude of $\rho_{21}$ depends on the phonon mean free 
path $\ell_{\rm ph}$ which appears in the phonon propagator.  
In Ref. \onlinecite{bonsager98} it was shown that for 
$\ell_{\rm ph}$ larger than a critical value $\ell_{\rm crit}$, the 
transresistivity is enhanced by to a collective mode (i.e. a 
zero in the real part of the screening function near $q=2k_F$).  
At zero magnetic field $\ell_{\rm crit}\sim 200\ \mu$m for 
$n=1.5\times 10^{15}\ {\rm m}^{-2}$.

The view taken in this paper is that frictional drag is
sufficiently well understood that it can be used to probe,
at least at a qualitative level,
the wavevector and frequency dependent polarization functions
of the coupled 2D electron layers.  Just such a probe is
urgently needed to test composite fermion theory of the 
half-filled Landau level.

\subsection{Coulomb drag at $\nu=1/2$}

Sakhi,\cite{sakhi97}, Ussishkin and Stern,\cite{ussishkin97} and
Kim and Millis\cite{kim96}, using Chern-Simons RPA theory,
independently predicted that at $\nu=1/2$ the low-temperature
transresistivity $\rho_{21}$ would be greatly enhanced compared to its
$B=0$ value.  In these theories $\rho_{21}\propto (T/d)^{4/3}$ for
$T\rightarrow 0$.  This property follows from the form of the
Chern-Simons RPA polarization function in the limits
$q\ll k_F$ and $\omega\ll v_Fq$ (see Appendix \ref{polapp}).
Subsequent experiments by Lilly {\it et al.},\cite{lilly98}
did demonstrate increased drag which vanished less rapidly
than $T^2$ with declining temperature.  The experimental results
are consistent with a $T^{4/3}$ law, except below $\sim 0.5\ {\rm K}$
where the temperature dependence slows further and the drag
value becomes sample specific.

In Chern-Simons RPA theory, the Coulomb drag enhancement 
is due to the diffusive 
character of composite-fermion density fluctuations at
long wavelength and low frequency.  
In Fig. \ref{coul} we plot the Coulomb contribution to $\rho_{21}$
(evaluated by numerical integration
of Eq. (\ref{rho21})) as a function of $d$ for different temperatures.
The plot shows that for $T\gtrsim 0.1$ K, the transresistivity falls of
approximately as $d^{-2}$, significantly slower than the
$d^{-4}$ behavior at $B=0$, but still fast enough to guarantee that
phonon-mediated drag will dominate for widely separated layers.
The plots in Fig. \ref{coul} were calculated using the
Chern-Simons RPA with an effective mass $m^*=m_{\rm e}$.
In Fig. \ref{massdep} we illustrate the
dependence of Coulomb drag on the choice of the effective mass,
and the difference between RPA and MRPA predictions.
Phonon mediated drag which is dominated by short wavelengths is 
not enhanced by the diffusion-like pole in the polarization 
function.\cite{khveshchenko} It will, however, as we discuss below, be 
enhanced over the corresponding zero-field result.  A numerical comparison 
shows that, generally, phonon-mediated drag will dominate over Coulomb drag for 
$d\gtrsim 200$ nm.


\section{Phonon mediated drag at $\nu=1/2$}\label{phonondrag}

Two qualitative aspects of the comparison between phonon-mediated drag
at $\nu=1/2$ and at $B=0$ can be anticipated in advance of any detailed calculation.
Firstly, since the polarization function
is proportional to the effective mass, one would expect the
phonon-mediated transresistivity to be larger at $\nu=1/2$ by a factor of approximately
$(m^*/m_b)^2\sim 100$.  Notice that this enhancement is weaker than
for Coulomb drag which is enhanced by a factor of 
approximately 1000 (independent of mass for $T\rightarrow 0$). 
 Secondly, the value of
$T_{\rm peak}$ should be larger than at $B=0$ for several reasons.
The Fermi wavevector is increased by a factor
of $\sqrt{2}$ if, as we assume here, the electron system is completely
spin polarized.  
The smaller Fermi energy and Fermi velocity, which
go along with the larger effective mass, should also increase
$T_{\rm peak}$.
For larger $m^*$, the ratio of the phonon velocity to the Fermi
velocity approaches unity, and the most important drag contributions will then
come from energies near the Fermi energy.  Correspondingly, within
the particle-hole continuum, the largest possible momentum transfer
(which is given by $q=2k_F(1+c_{\lambda}/v_F)$) is larger at
$\nu=1/2$ than for $B=0$ ($c_{\lambda}\ll v_F$ for $B=0$).
Furthermore, since $T_{\rm peak}$ is close to the Fermi temperature
$T_F$, it is necessary to evaluate the
polarization functions at finite temperature. (At finite $T$ the
Fermi gas polarization function can no longer be evaluated
analytically.  See Appendix \ref{polapp} for
a detailed discussion of its numerical evaluation.)
When evaluated at finite temperature, the polarization function has
finite contributions from outside the $T=0$ particle-hole continuum (see Fig.
\ref{phasespace}).  This additional effect suggests that a
{\it further} increase of $T_{\rm peak}(\nu=1/2)/T_{\rm peak}(B=0)$ 
which should therefore 
be larger than $\sqrt{2} (1+c_{\lambda}/v_F)$.  Indeed our numerical
calculations confirm this expectation which is not, however, in
agreement with experiment.

Zelakiewicz {\it et al.}\cite{zelakiewicz} have measured the transresistivity
in samples with $d=2600$ \AA\ and $d=5200$ \AA\ where they find no evidence of
a contribution from the Coulomb interaction.  The transresistivity was indeed
found to have the same qualitative features as at $B=0$, except that the
transresistivity is enhanced by a factor of $\sim 200$.  The scaled
transresistivity $\rho_{21}/T^2$ does have a peak as a function of temperature
whose location scales with electron density as
$\sqrt{n}$, just as in the $B=0$ case.
However, the value of $T_{\rm peak}$ was found to
be slightly smaller at $\nu=1/2$ than at $B=0$.
The detailed calculations we present below thus tend to indicate that
Chern-Simons composite fermion theory overestimates
electron polarizability at wavevectors $\sim 2 k_F$, 
as anticipated by the authors of Ref. \onlinecite{zelakiewicz}.
Our calculations are similar to ones performed earlier\cite{bonsager98} at
$B=0$; the most troublesome complication is the requirement that
the free-particle polarization function be evaluated
at finite temperatures.  This aspect of the calculation is explained in
detail in Appendix \ref{polapp}.

Figs. \ref{mrpa0067}, \ref{mrpa04},
and \ref{mrpa10} show numerical evaluations of $\rho_{21}/T^2$ as a
function of $T$ for $m^*=0.067 m_e$, $m^*=0.4m_e$ and $m^*=m_e$ respectively.
Calculations at intermediate values of $m^*$ interpolate smoothly
between the trends illustrated in these three figures.
The striking differences we find in the temperature and density dependences for
the different values of $m^*$ are due primarily to the different values of the
Fermi temperature compared to the energy of an acoustic phonon with a
wavelength comparable to the distance between electrons.
We concentrate first on the property which is most prominent in
the experimental data, as conventionally presented, $T_{\rm peak}$.
We discuss its dependence on
density and the phenomenological mass parameter $m^*$.
At zero magnetic field $T_{\rm peak} = 2.6\ {\rm K}$ in a calculation 
which uses the same parameters as in Fig. \ref{mrpa0067}.
Fig. \ref{mrpa0067} shows results obtained when the band mass
is used for $m^*$ so that differences from the $B=0$ case are
due only to spin-polarization and the introduction of the magnetic local field.
The local field increases the relative importance of small angle
scattering and decreases $T_{\rm peak}$ below the naively expected value
$\sqrt{2}\times 2.6$ K.  This tendency of the local field is closely
connected to the modification which changes
the temperature dependence from $T^2$ to $T^{4/3}$ in the Coulomb drag case.
Even with this effect, however, $T_{\rm peak}$ is safely
above its zero field value, in disagreement with the experimental
findings of Ref. \onlinecite{zelakiewicz}.
For this mass, $T_{\rm peak}$ increases with density approximately as $\sqrt{n}$ 
as in the $B=0$ case.\cite{bonsager98}  The origin of this behavior is
simply that the highest momentum transfer within the
particle-hole continuum is proportional to $\sqrt{n}$ and that
the Fermi temperature for these values of $m^*$ is sufficiently
high compared to $T_{\rm peak}$ that Im$\Pi(q,\omega)$ has
negligible weight outside the particle-hole continuum.

In Figs. \ref{mrpa04} and \ref{mrpa10} we see that $T_{\rm peak}$
is still larger at the larger values of $m^*$ required to fit other data
in the quantum Hall regime.
This is in accord with the considerations outlined above.
With increasing $m^*$ the Fermi temperature decreases and
momentum transfers greater than $2k_F$ receive more
weight.  For higher values of the mass, it is interesting to see that the
value of $T_{\rm peak}$ has a more complex density dependence,
reflecting a competition between two effects.
As the density increases, the maximum possible momentum transfer
increases, which tends to increase $T_{\rm peak}$.  At the same
time, however, the Fermi temperature increases which leads to a
smaller contribution from outside the particle-hole continuum.
This tends to decrease $T_{\rm peak}$.
Because of this competition, the density-dependence of 
$T_{\rm peak}(n)$ changes from increasing at  
$m^*=0.067m_e$, to decreasing at $m^*=0.4m_e$, to
approximately constant for $m^*>0.4m_e$.
As for the magnitude of the transresistivity, we see that
for a phonon mean free path $\ell_{\rm ph}=100\ \mu {\rm m}$,
we would have to choose an effective mass in the neighborhood of
$m^*=0.3m_{\rm e}$ in order to match the experimental result
that $\rho_{21}/T_{\rm peak}^2\simeq 200\ {\rm m}\Omega/{\rm K}^2$.
(For this choice of parameters , the numerical results
need to be multiplied by a
factor of 5.7 to match the experimental results at
zero magnetic field\cite{noh99}).

Phonon mediated drag might not, however, be a good way of determining the
effective mass of Composite Fermions.  
The critical phonon mean free path $\ell_{\rm crit}$
itself depends on $m^*$ ($\ell_{\rm crit}\propto 1/(m^*)^2$) 
and will thus be different for
$B=0$ and $\nu=1/2$.  In Fig. \ref{lphdep} we plot the
transresistivity as a function of $\ell_{\rm ph}$ for different
values of $m^*$.  At $\ell_{\rm ph}\simeq\ell_{\rm crit}$ a
long-lived collective mode develops, leading to an enhancement of $\rho_{21}$
before it saturates at $\ell_{\rm ph}\gg\ell_{\rm crit}$.
The qualitative distance dependence is rather similar
to the zero magnetic field case (see Ref. \onlinecite{bonsager98}):
For $\ell_{\rm ph}\ll\ell_{\rm crit}$, the transresistivity depends
logarithmically on $d/\ell_{\rm ph}$, i.e. 
$\rho_{21}\propto \ln (Ad/\ell_{\rm ph})$ for $d\ll\ell_{\rm ph}/A$ 
where $A$ is a constant of order $2k_FL$.  For $d\gg\ell_{\rm ph}/A$ 
the transresistivity falls off exponentially, 
$\rho_{21}\propto\exp(-Ad/\ell_{\rm ph})/d$.  
If, on the other hand, the phonon mean free path is larger than the 
critical value $\ell_{\rm crit}$, the distance dependence is more 
complicated.

\section{Discussion}\label{conclusions}

In this paper we describe the dependence of phonon mediated drag 
at $\nu=1/2$ on system parameters and composite fermion mass 
predicted by the Chern-Simons MRPA theory.  We have focused on the 
characteristic temperatures $T_{\rm peak}$ at which the 
scaled drag resistivity $\rho_{21} (T) / T^2$ reaches its peak. 
Crudely we expect $k_B T_{\rm peak}$ to equal 
the phonon energy at the largest wavevector for which the 
two-dimensional electron systems have substantial charge fluctuations.
Because of the sharp particle-hole continuum of Chern-Simons 
composite fermion theory, the maximum momentum transfer is close to 
$2k^{\rm cf}_F$, where $k^{\rm cf}_F$ is the Fermi wavevector of the
composite fermion Fermi sea.  These considerations lead to    
$T_{\rm peak} \sim {\hbar} c_{\lambda} 2 k^{\rm cf}_F/k_B$.
Since $k^{cf}_F = \sqrt{2} k_F$ for completely spin polarized
composite fermions, $T_{\rm peak}$ at $\nu=1/2$ should be larger than 
$T_{\rm peak}$ at $B=0$, in disagreement with experiment.
Our detailed calculations confirm that, at a qualitative level,
this is indeed the prediction of Chern-Simons composite fermion theory. 
 
At a quantitative level, the change of Fermi radius is not the 
only difference which arises in comparing phonon-mediated drag
at $B=0$ and at $\nu=1/2$.
One source of complication is the strong dependence of the drag
resistivity on the effective mass $m^*$ of the composite fermions.
The large effective mass of composite fermions leads to a substantially
smaller Fermi energy and Fermi velocity. 
This in turn leads to larger maximum momentum transfer,
$q = 2k^{\rm cf}_F (1 + c_{\lambda}/v_F)$, since  
$c_{\lambda}$ is then comparable to $v_F$.
In addition, contributions to drag from outside the 
$T=0$ particle-hole continuum increase.
On the other hand, the magnetic local field correction increases
the relative importance of small angle scattering and tends to 
decrease $T_{\rm peak}$.
Because of these mutually competing effects, the position of
the maximum is sometimes larger than value expected on the basis
of naive considerations and quoted above.  For example, when $m^*$ is
comparable to $m_b$, the Fermi temperature effect is relatively
small and the magnetic local field effect leads to a 
$T_{\rm peak}$ smaller than the naive value.
Nevertheless, the position of $T_{\rm peak}$ at $\nu=1/2$ is 
{\em always} well above that of $B=0$ case.

We now present a detailed summary of the comparison between 
our theoretical results and experiment \cite{zelakiewicz}.
The following aspects of the experimental data \cite{zelakiewicz} 
are consistent with theoretical results. 
\begin{enumerate}

\item The magnitude of the phonon mediated drag at $\nu=1/2$ is 
about 200 times larger than at $B=0$. In the theory, the
factor of a few hundred comes from the enhanced density of 
states, $m^*/2\pi\hbar^2$, of composite fermions compared to the bare 
density of states, $m_b/2\pi\hbar^2$, of electrons at $B=0$. 
$m^*/m_b \sim 10$ according to
numerical and theoretical estimations \cite{hlr}. 
The presence of two layers leads to $(m^*/m_b)^2 \sim 100$
fold enhancement in the drag rate.

\item When the phonon mean free path of $\ell_{ph} = 100 \ \mu m$ is
chosen consistently for $B=0$ and $\nu=1/2$, the experimental
data can be fit by choosing $m^*$ around $0.3 m_e$ which is
consistent with previous numerical and theoretical estimations 
\cite{hlr}.

\item There is a well defined maximum of $\rho_{21}/T^2$,
suggesting that $\nu=1/2$ systems have a fairly sharp wavevector
cutoff for low-energy charge fluctuation, as predicted by 
Chern-Simons composite fermion theory.

\end{enumerate}
On the other hand, the following experimental results do not
agree with theoretical predictions.

\begin{enumerate}

\item Experimentally, $T_{\rm peak}$ at $\nu=1/2$ is always smaller 
than that of $B=0$. Theoretically it is always larger, even when
unrealistically small values are used for the composite fermion
effective mass.

\item Experimentally, $T_{\rm peak}$ is 
proportional to $\sqrt{n}$.  Theoretically, this simple behavior 
is found only when unrealistically small values are used for the 
composite fermion effective mass.

\end{enumerate}

We now discuss some possible sources of these discrepancies.

{\it Incomplete spin polarization}: 
In our calculations, we assumed complete spin polarization.
If the spins are only partially polarized, the Fermi sea of the majority spin
would be smaller than in the fully polarized case, leading  
to a smaller maximum momentum transfer and a smaller $T_{\rm peak}$.
Recently the spin polarization of $\nu=1/2$ state was measured by 
NMR \cite{melinte,dementyev}.
In the experiment by S. Melinte et al. \cite{melinte}, the electron 
density, $n = 1.4 \times 10^{11} cm^{-2}$, of their sample M242 is similar 
to the density, $n = 1.5 \times 10^{11} cm^{-2}$, of the sample studied 
in the drag experiment of S. Zelakiewicz et al. \cite{zelakiewicz}.
In the NMR experiment \cite{melinte},
$\nu=1/2$ is reached when the external magnetic field is $B=11.4$ T,
compared to the field strength $B = 12.82$ T for $\nu=1/2$ 
in the drag experiment \cite{zelakiewicz}.
This NMR experiment can therefore be used to get a reasonable
estimate of the $\nu = 1/2$ spin polarization in the drag experiment. 
We conclude that the spin polarization in the $\nu=1/2$ drag 
experiment at $1-2$ K was about $70 \%$ of the full
polarization.  This would imply that majority-spin Fermi wavevector was 
$\sqrt{1.7} k_F \approx 1.3 k_F$ and that of the minority spin
Fermi wavevector is $\sqrt{0.3} k_F \approx 0.55 k_F$.
This reduction in the majority spin-wavevector 
($1.3 k_F$ compared to $\sqrt{2} k_F$) is not sufficient 
to explain the low $T_{\rm peak}$ values seen in the experiments.
In our theory, we would not expect a large contribution to 
drag from the minority spins, and in any event 
the Fermi radius $0.55 k_F$ is too small for it to be
associated with the observed value of $T_{\rm peak}$ 
slightly below that of $B=0$ case.  We conclude that 
incomplete spin polarization alone cannot explain the discrepancy
between theory and experiment.
 
{\it Breakdown of (M)RPA at large wavevectors}:
This experimental result may signal that the (M)RPA is not adequate 
to describe the large wavevector response of a $\nu=1/2$ system.
It is generally accepted that Ward identities \cite{kim94} related 
to conservation laws, limit beyond (M)RPA contributions to 
density-density and current-current response functions 
in the long wavevectors and low energy limit.  However, it has been also
shown that response functions at large wavevectors, in particular near $2k_F$,
may be modified at low energies by singular vertex corrections \cite{ioffe94}.
Even though these singular corrections appear at low energies at $2k_F$,
it is certainly possible that these singular corrections persist
to higher energy scales comparable to $T_{\rm peak}$ or 
${\hbar} c_{\lambda} 2 k^{\rm cf}_F/k_B$. If this is true, it may
be necessary to get beyond the (M)RPA to obtain an accurate description
of phonon-mediated drag.

{\it Dipolar Composite Fermions}:  
As discussed in the introduction, there exist at present two 
descriptions of $\nu = 1/2$ composite fermions.  It has been 
established that these two descriptions
are equivalent in the low energy and long wavelength limits.
However, these two descriptions may lead to quite different
predictions at large wavevectors because 
dipolar composite fermions have a finite size which can be 
comparable to $k_F^{-1}$.
Indeed, it has been observed that the equivalence of two approaches
may break down at higher energies even in the long
wavelength limit \cite{shankar99}. 
The singular behavior of the response 
functions near $2k_F$ mentioned above in the Chern-Simons theory
approach may be a signature of the breakdown of the theory at
large wavevector scales and a proper description of the system
at large wavevectors may require the fully lowest Landau level
dipolar composite-fermion theory. 
Calculations of the relevant response functions and of the  
drag resistivity in the dipolar composite fermion approach
are in progress \cite{bkm}.

In conclusion, the Chern-Simons theory of composite fermions 
overestimates charge fluctuations at the Fermi wavelength scale
and a proper description of the system at large wavevectors 
may have to take into account the extended nature of dipolar
composite fermions.

\section{Acknowledgements}

This work was initiated at the Institute for Theoretical Physics in University
of California at Santa Barbara (NFS grant PHY9407194), and was 
further supported by the NSF under grants DMR-9714055 (M.C.B and A.H.M) and 
DMR-9983783 (Y.B.K.), by the Danish Research Academy (M.C.B), and by
the A. P. Sloan Foundation (Y.B.K.).
The authors are grateful T. J. Gramila, B. Y.-K. Hu,
and S. Zelakiewicz, for informative and stimulating interactions.
Y.B.K. would like to thank also Isaac Newton Institute at the University of 
Cambridge for the hospitality.


\appendix

\section{Polarization functions}\label{polapp}

In this appendix we summarize the techniques we use to evaluate
the free-fermion polarization functions which appear in Eq. (\ref{rho21}) at finite
temperatures and, for completeness, present explicit  
expressions for the (Modified) Random Phase Approximation 
polarization functions.

The electron polarization function $\Pi^{\rm e}(q,\omega)$ expresses the linear
response of an electron system to the total electromagnetic field.
For a two-dimensional system (in the $x$-$y$ plane) and
a magnetic induction restricted to the $z$-direction, ${\bf B}=B\hat{z}$,
it is sufficient to consider the number density $n$ and the $y$-component of the
number current density $j_y$ ($j_x$ is fixed by particle conservation).
In a 2-vector notation, the polarization function is defined by
\begin{equation}
j^{\alpha}(q,\omega)=ec^2 \Pi^{{\rm e},\alpha\beta}(q,\omega)
A_{\rm total}^{\beta}(q,\omega),
\end{equation}
where ${\bf j}=(cn,j_y)$ and ${\bf A}=(\Phi/c,A_y)$.  The scalar potential
$\Phi$ and vector potential ${\bf A}$ are related to the electric field
${\bf E}$ and the magnetic induction ${\bf B}$ is given by
${\bf E}=-{\bf \nabla}\Phi$ and ${\bf B}={\bf \nabla}\times {\bf A}$
with ${\bf \nabla}\cdot {\bf A}=0$.  This choice is known as the Coulomb gauge.
The total field is the sum of the external field and the field induced
by the current and charge response of the system.

The tensor $\Pi^{\rm e}(q,\omega)$ has four components
$\Pi^{\rm e}_{00}$, $\Pi^{\rm e}_{01}$, $\Pi^{\rm e}_{10}$, and $\Pi^{\rm e}_{11}$,
which are related by linear response theory
to density-density, density-current, current-density, and current-current
correlation functions, respectively.
For the drag calculations we are only interested in the density-density
response $\Pi^{\rm e}_{00}$ which is called $\Pi(q,\omega)$ in the main text.

$\Pi^{\rm e}(q,\omega)$ is known only approximately, even at $B=0$.
In the Random Phase Approximation, the polarization function correlation functions are
approximated by their non-interacting fermion forms.
At $B=0$ we thus use $\Pi^{\rm e}=2\Pi^{(0)}_{00}$ with
\begin{equation}
\Pi^{(0)}_{00}=\frac{1}{A}\sum_{\bf k}
\frac{n_F(\xi_{\bf k+q})-n_F(\xi_{\bf k})}
{\hbar\omega-\xi_{\bf k+q}+\xi_{\bf k}+i\eta}.
\end{equation}
At zero temperature the wavevector integral can be evaluated analytically:\cite{stern67}
\begin{eqnarray}
{\rm Re}\Pi^{(0)}_{00}&=&\frac{g_0}{2z}\left( 2z-C_{-} [(z-u)^2-1]^{1/2}\right.
\nonumber\\
&&
\left. -C_{+} [(z+u)^2-1]^{1/2} \right)
\end{eqnarray}
\begin{eqnarray}
{\rm Im}\Pi^{(0)}_{00}&=&\frac{g_0}{2z}
\left( D_{-} [1-(z-u)^2]^{1/2}\right.\nonumber\\
&&\left.
-D_{+} [1-(z+u)^2]^{1/2}\right)
\end{eqnarray}
where $g_0=m_b/2\pi\hbar^2$, $z=q/2k_F$, $u=\omega/qv_F$, $m_b$ is the
band mass, $v_F$ is the Fermi velocity, and
\begin{equation}
C_{\pm}=\left\{
\begin{array}{ll}
\frac{z\pm u}{|z\pm u|}  &  ,|z\pm u|>1  \\
0                        &  ,|z\pm u|<1
\end{array}
\right.
\end{equation}
\begin{equation}
D_{\pm}=\left\{
\begin{array}{ll}
0  &  ,|z\pm u|>1  \\
1  &  ,|z\pm u|<1
\end{array}
\right.
\end{equation}

In the fractional quantum Hall regime, electron-electron interactions
cannot be ignored.  In Chern-Simons composite fermion theory
interactions enter only through the local field produced by attached
flux quanta.  The composite fermion polarization function is
defined by
\begin{equation}
j^{\alpha}(q,\omega)=ec^2 \Pi_{\rm CF}^{\alpha\beta}(q,\omega)
A_{\rm CF,total}^{\beta}(q,\omega),
\end{equation}
where ${\bf A}_{\rm CF,total}$ is the total electromagnetic field
seen by the Composite Fermions which includes the Chern-Simons field.
In the RPA (MRPA) one approximates $\Pi_{\rm CF}$ by $\Pi^{(0)}$ and relates
$\Pi^{\rm e}$ to $\Pi_{\rm CF}$ by\cite{hlr,simon93}
\begin{equation}
[\Pi^{\rm e}]^{-1}=[\Pi_{\rm CF}]^{-1}+{\bf C}+{\bf F}
\end{equation}
with
\begin{equation}
{\bf C}=\frac{ec\Phi_0\tilde{\phi}}{q}\left[
\begin{array}{cc}
0  &  i \\
-i &  0
\end{array}
\right]
\end{equation}
and
\begin{equation}
{\bf F}=\frac{\Delta m}{n_0}\left[
\begin{array}{cc}
(\frac{\omega}{q})^2  &  0 \\
0 &  c^2
\end{array}
\right]
\end{equation}
Here $\Delta m=m^*-m_b$ where $m^*$ it the effective mass, and $n_0$ is the
equilibrium electron number density, $\Phi_0=2\pi\hbar/e$, and $\tilde{\phi}=2$ 
at half filling.  
The difference between RPA and MRPA lies in the matrix ${\bf F}$ which
is set to zero for the RPA.  Notice that MRPA reduces to RPA if the effective
mass is chosen to be equal to the band mass.
Solving for $\Pi^{\rm e}_{00}$ we get (since $\Pi_{01}^{(0)}=\Pi_{10}^{(0)}=0$)
\begin{equation}\label{mrpaexpr}
\Pi^{\rm e}_{00}=\frac{\Pi^{(0)}_{00}}
{1+\frac{\Delta m}{n_0}\left(\frac{\omega}{q}
\right)^2\Pi^{(0)}_{00}-\left(\frac{ec\Phi_0\tilde{\phi}}{q}\right)^2
\Pi^{(0)}_{00}\Pi^{(0)}_{11}
\frac{n_0}{n_0+c^2\Delta m\Pi^{(0)}_{11}}}
\end{equation}
In Eq. ({\ref{mrpaexpr}) the band mass in the expression for $\Pi^{(0)}_{00}$
is replaced by the effective mass.  $\Pi_{11}^{(0)}$ is related to the 
current-current correlation function and is given by
\begin{equation}
\Pi^{(0)}_{11}=\frac{-n_0}{m^*c^2}+\frac{1}{A}\sum_{\bf k}
\left( \frac{\hbar k_y}{m^*c} \right)^2
\frac{n_F(\xi_{\bf k+q})-n_F(\xi_{\bf k})}{\hbar\omega-\xi_{\bf k+q}+\xi_{\bf k}+i\eta}
\end{equation}
which at zero temperature is 
\begin{eqnarray}
{\rm Re}\Pi^{(0)}_{11}&=&\frac{-n_0}{m^*c^2}+(\frac{\hbar k_F}{m^*c})^2\frac{g^*_0}{6z}
\left(   3z+C_{+}[(z+u)^2-1]^{3/2}\right. \nonumber\\
&& \left. +C_{-}[(z-u)^2-1]^{3/2} -(z+u)^3-(z-u)^3 \right)\nonumber\\
\end{eqnarray}
\begin{eqnarray}
{\rm Im}\Pi^{(0)}_{11}&=&(\frac{\hbar k_F}{m^*c})^2\frac{g^*_0}{6z}
\left( D_{-}[1-(z-u)^2]^{3/2}\right.\nonumber\\
&&\left.
-D_{+}[1-(z+u)^2]^{3/2} \right)
\end{eqnarray}
with $g^*_0=m^*/2\pi\hbar^2$.

At finite temperature, the expressions for $\Pi_{00}^{(0)}$ and 
$\Pi_{11}^{(0)}$ must be evaluated numerically.  We have done this by 
evaluating the angular integral analytically and 
expressing the functions in terms of Fermi-Dirac Integrals which are 
defined according to 
\begin{equation}
{\cal F}_j(x,b)=\frac{1}{\Gamma(j+1)}\int_b^{\infty}
\frac{t^j}{1+\exp(t-x)}dt.
\end{equation}
We find
\begin{eqnarray}
(g^*)^{-1}{\rm Re}\Pi_{00}^{(0)}&=&1-
\frac{\sqrt{\pi}\tilde{T}^{1/2}}{4z}\left(
{\cal F}_{-1/2}(\frac{\Omega_{+}^2-\tilde{\mu}}{\tilde{T}},0)\right.\nonumber\\
&&\left.
-
{\cal F}_{-1/2}(\frac{\Omega_{+}^2-\tilde{\mu}}{\tilde{T}},
\frac{\Omega_{+}^2}{\tilde{T}})\right)\nonumber\\
&-&{\rm sgn}(\Omega_{-})\frac{\sqrt{\pi}\tilde{T}^{1/2}}{4z}\left(
{\cal F}_{-1/2}(\frac{\Omega_{-}^2-\tilde{\mu}}{\tilde{T}},0)\right.\nonumber\\
&&\left.
-
{\cal F}_{-1/2}(\frac{\Omega_{-}^2-\tilde{\mu}}{\tilde{T}},
\frac{\Omega_{-}^2}{\tilde{T}})\right)
\end{eqnarray}
\begin{eqnarray}
(g^*)^{-1}{\rm Im}\Pi_{00}^{(0)} &=&
\frac{\sqrt{\pi}\tilde{T}^{1/2}}{4z}\left(
{\cal F}_{-1/2}(\frac{\tilde{\mu}-\Omega_{-}^2}{\tilde{T}},0)\right.\nonumber\\
&&\left.
-
{\cal F}_{-1/2}(\frac{\tilde{\mu}-\Omega_{+}^2}{\tilde{T}},0)\right)
\end{eqnarray}
\begin{eqnarray}
(\frac{m^*c}{\hbar k_F})^2(g_0^*)^{-1}{\rm Re}\Pi_{11}^{(0)}&=&-u^2-z^2/3\nonumber\\
&+&
\frac{\sqrt{\pi}\tilde{T}^{3/2}}{8z}\left(
{\cal F}_{1/2}(\frac{\Omega_{+}^2-\tilde{\mu}}{\tilde{T}},0)\right.\nonumber\\
&&\left.
-
{\cal F}_{1/2}(\frac{\Omega_{+}^2-\tilde{\mu}}{\tilde{T}},\frac{\Omega_{+}^2}{\tilde{T}})
\right)\nonumber\\
&+&{\rm sgn}(\Omega_{-})\frac{\sqrt{\pi}\tilde{T}^{3/2}}{8z}\left(
{\cal F}_{1/2}(\frac{\Omega_{-}^2-\tilde{\mu}}{\tilde{T}},0)\right.\nonumber\\
&&\left.
-
{\cal F}_{1/2}(\frac{\Omega_{-}^2-\tilde{\mu}}{\tilde{T}},\frac{\Omega_{-}^2}{\tilde{T}})
\right)
\end{eqnarray}
\begin{eqnarray}
(\frac{m^*c}{\hbar k_F})^2(g_0^*)^{-1}{\rm Im}\Pi_{11}^{(0)} &=&
\frac{\sqrt{\pi}\tilde{T}^{3/2}}{8z}\left(
{\cal F}_{1/2}(\frac{\tilde{\mu}-\Omega_{-}^2}{\tilde{T}},0)\right.\nonumber\\
&&\left.
-
{\cal F}_{1/2}(\frac{\tilde{\mu}-\Omega_{+}^2}{\tilde{T}},0)\right)
\end{eqnarray}
Here $\Omega_{\pm}=z\pm u$, and we have defined the dimensionless 
quantities $\tilde{\mu}=\mu/E_F$ and $\tilde{T}=k_BT/E_F$ 
where $\mu$ is the chemical potential and 
$E_F$ is the Fermi energy.  For parabolic bands 
$\tilde{\mu}/\tilde{T}=\ln(e^{1/\tilde{T}}-1)$.
For evaluating the Fermi-Dirac Integrals we have used efficient 
algorithms that are publicly available.\cite{fermidirac}



\begin{figure}
\epsfxsize=6cm
\rotate[r]{\epsfbox{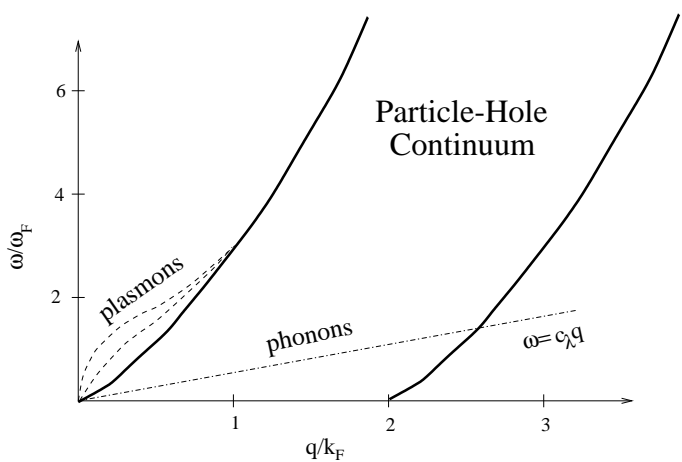}}
\vspace{0.5cm}
\caption{
A sketch of the $(q,\omega)$-space on which the integrand of Eq. (\ref{rho21}) is
defined.  At zero temperature Im$\Pi(q,\omega)$ is zero outside the particle-hole
continuum.  However, at finite temperature the integrand will have finite weight
at the position of the plasmon poles indicated by the dashed lines.\cite{flensberg94}
(There are two plasmon poles whose exact position and strength depend on the interlayer
distance).
The dash-dotted line shows the phonon resonance.  Phonon mediated drag is dominated
by contributions from this resonance for as high as possible $q$ limited by
$\omega<T$.
}
\label{phasespace}
\end{figure}

\begin{figure}
\epsfxsize=6.5cm
\rotate[r]{\epsfbox{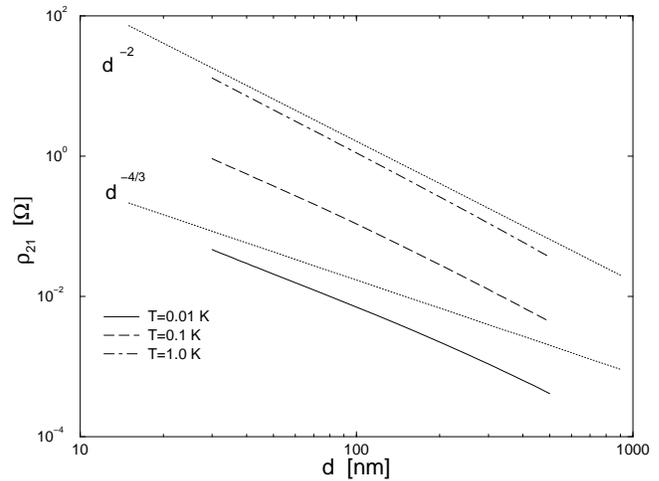}}
\vspace{0.5cm}
\caption{
The Coulomb contribution to the transresistivity calculated as a function
of distance between the layers
using the RPA with $m^*=m_{\rm e}$.  The two dotted lines indicate the
power laws $d^{-4/3}$ and $d^{-2}$, respectively. 
The calculations are based on Eq. (\ref{rho21}) and the RPA for composite 
fermions.  
}
\label{coul}
\end{figure}

\begin{figure}
\epsfxsize=6.5cm
\rotate[r]{\epsfbox{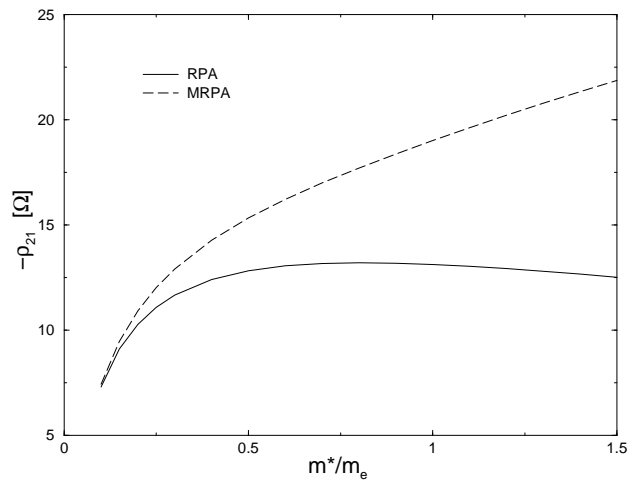}}
\vspace{0.5cm}
\caption{
The dependence on effective mass of the Coulomb contribution to the
transresistivity.  The temperature is 1 K (notice that there is no
mass dependence for $T\rightarrow 0$).  Other parameters are
$d=300$ \AA, $L=200$ \AA, and $n=1.5\times 10^{15}\ {\rm m}^{-2}$.
}
\label{massdep}
\end{figure}

\begin{figure}
\epsfxsize=6.5cm
\rotate[r]{\epsfbox{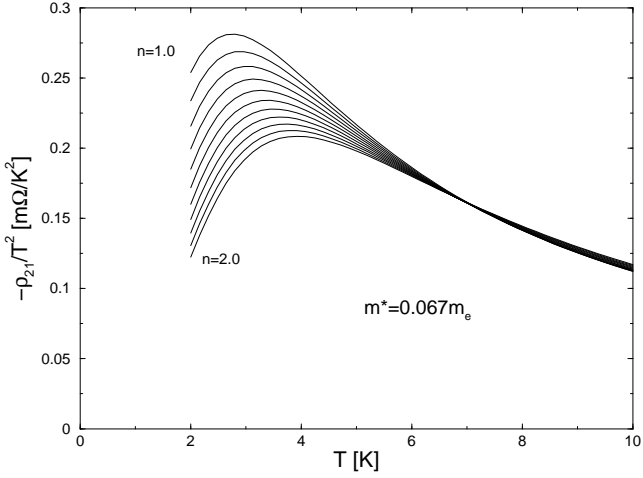}}
\vspace{0.5cm}
\caption{
Scaled transresistivity as a function of temperature for $m^*=m_b=0.067m_{\rm e}$.
The 11 curves are calculations for the density ranging from
$n=1.0\times 10^{15}\ {\rm m}^{-2}$ to $n=2.0\times 10^{15}\ {\rm m}^{-2}$
in increments of $n=0.1\times 10^{15}\ {\rm m}^{-2}$.
Other parameters are
$d=5000$ \AA, $L=200$ \AA, and $\ell_{\rm ph}=100\ \mu {\rm m}$.
For $n=1.5\times 10^{15}\ {\rm m}^{-2}$ the Fermi temperature is 62.3 K.
These calculations are based on the MRPA; corresponding calculations based on 
the RPA show almost identical results.    
}
\label{mrpa0067}
\end{figure}

\begin{figure}
\epsfxsize=6.5cm
\rotate[r]{\epsfbox{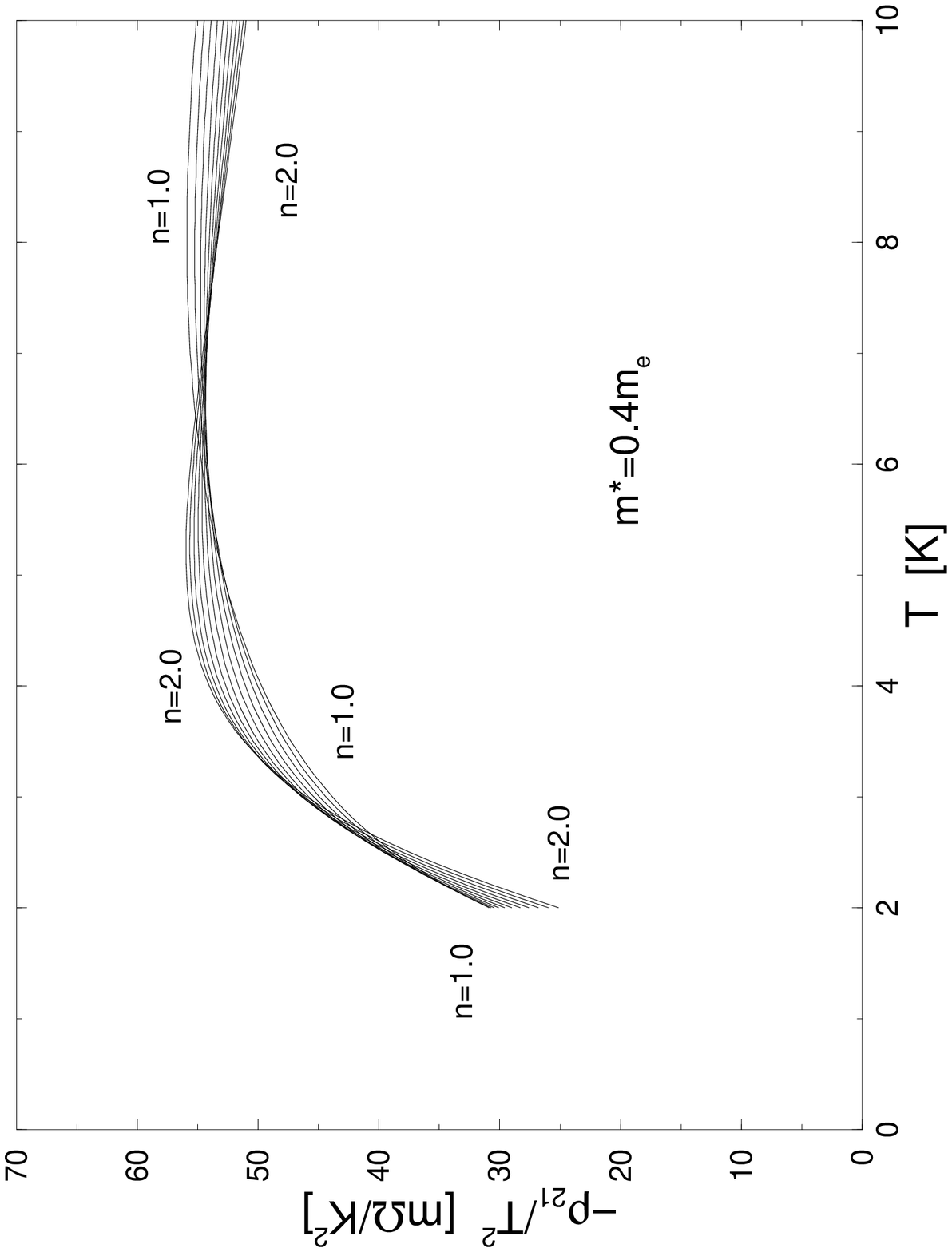}}
\vspace{0.5cm}
\caption{
The same as in Fig. \ref{mrpa0067} but with $m^*=0.4m_{\rm e}$.
For $n=1.5\times 10^{15}\ {\rm m}^{-2}$ the Fermi temperature is 10.4 K.
}
\label{mrpa04}
\end{figure}

\begin{figure}
\epsfxsize=6.5cm
\rotate[r]{\epsfbox{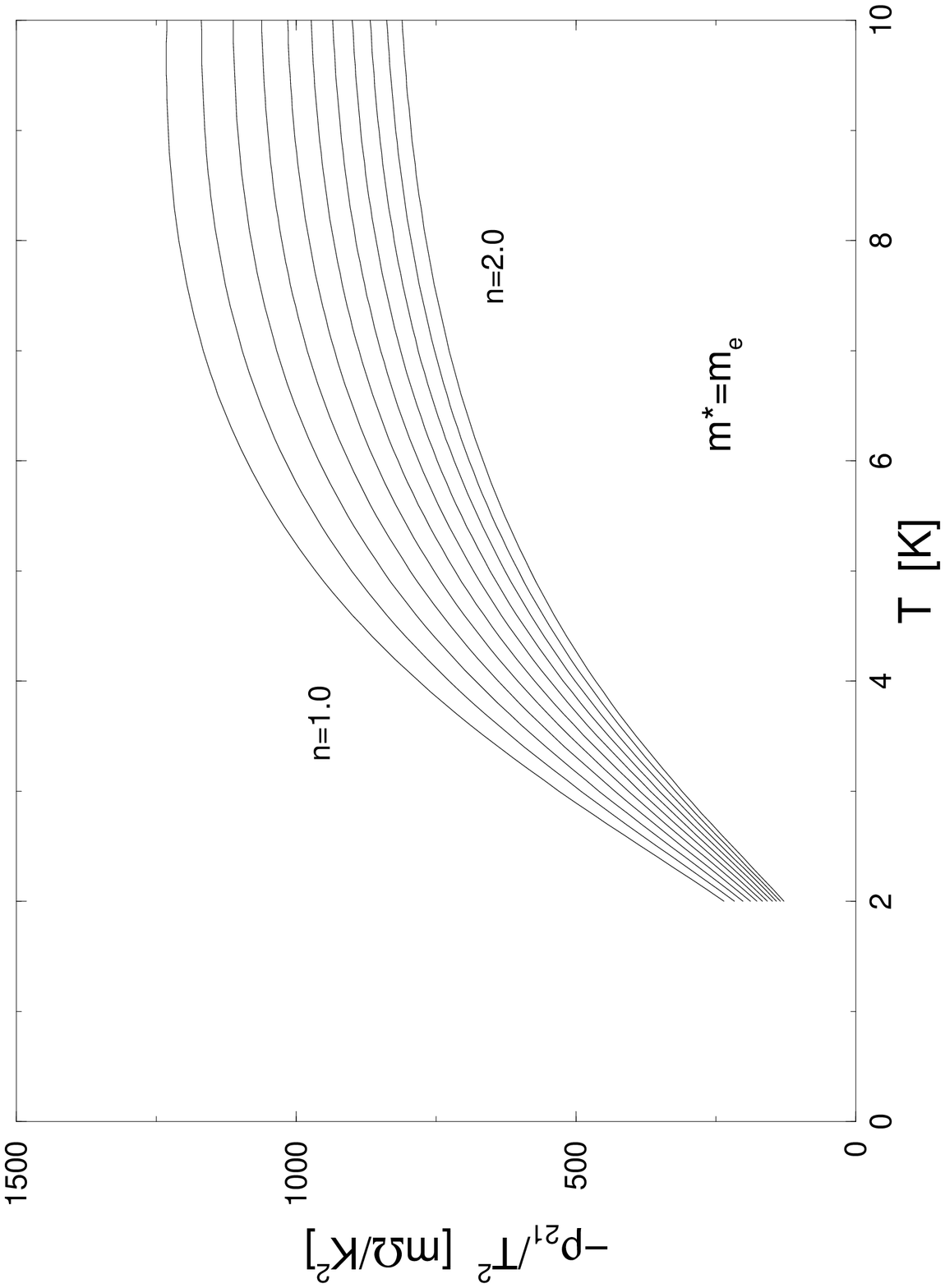}}
\vspace{0.5cm}
\caption{
The same as in Fig. \ref{mrpa0067} but with $m^*=m_{\rm e}$.
For $n=1.5\times 10^{15}\ {\rm m}^{-2}$ the Fermi temperature is 4.2 K.
}
\label{mrpa10}
\end{figure}

\begin{figure}
\epsfxsize=6.5cm
\rotate[r]{\epsfbox{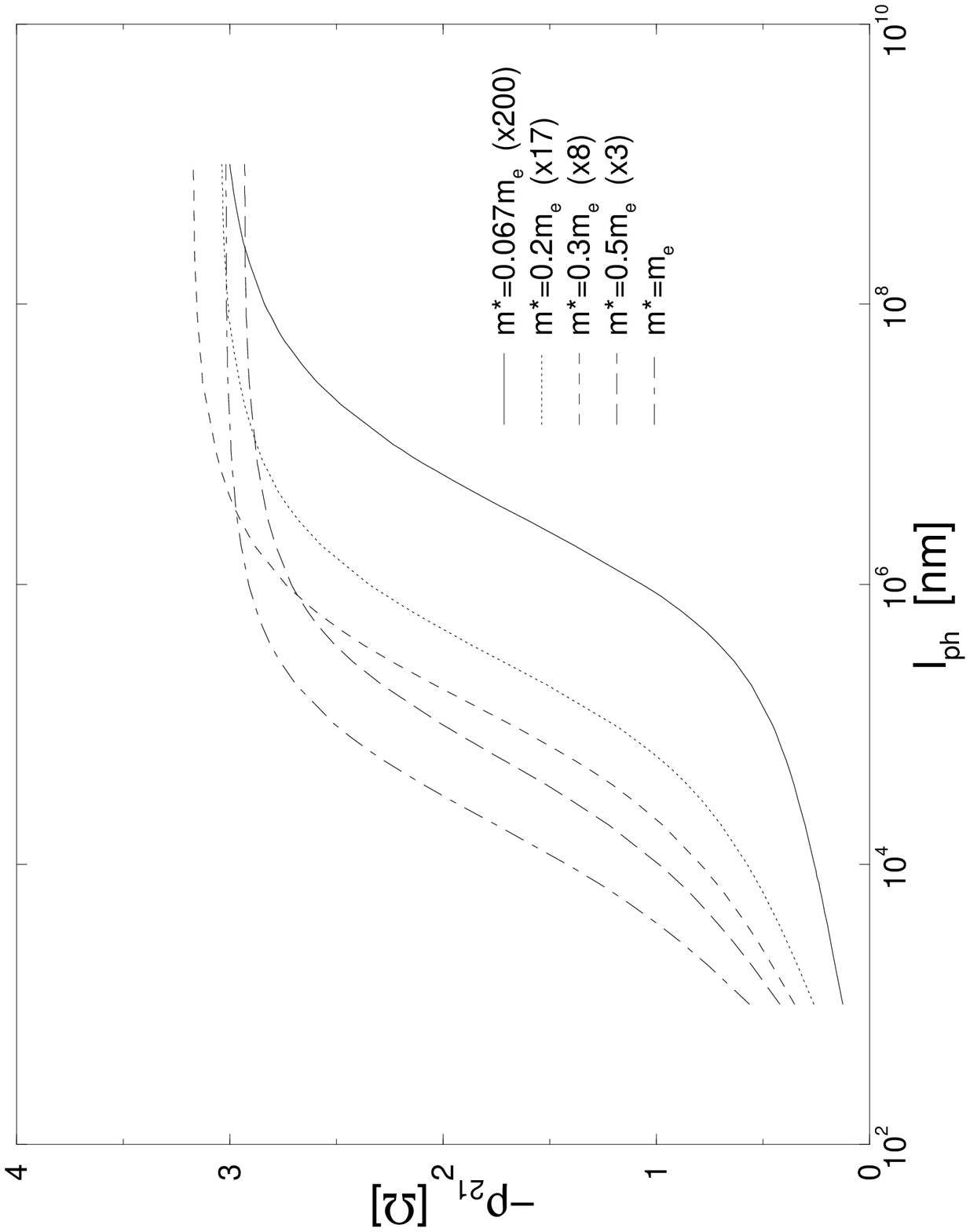}}
\vspace{0.5cm}
\caption{
The transresistivity as a function of phonon mean free path for different
values of the effective mass.  The plot shows how the critical phonon
mean free path $\ell_{\rm crit}$ decreases with increasing $m^*$.
Here, the temperature is 3 K and the distance is $d=500$ \AA.
}
\label{lphdep}
\end{figure}


\begin{thebibliography}{20}


\bibitem{stormer}
For a review, see H. L. Stormer, in {\it Perspectives in 
Quantum Hall Effects}, Edited by S. Das Sarma and 
A. Pinczuk (John Wiley and Sons, Inc., New York, 1997).


\bibitem{hlr} B. I. Halperin, P. A. Lee, and N. Read, Phys. Rev. B 
{\bf 47}, 7312 (1993).


\bibitem{jain} J. K. Jain, Phys. Rev. Lett. {\bf 63}, 199 (1989);
Adv. Phys. {\bf 41}, 105 (1992)).


\bibitem{kohn61} W. Kohn, Phys. Rev. {\bf 123}, 1242 (1961).


\bibitem{simon93}
S. H. Simon and B. I. Halperin, Phys. Rev. B {\bf 48}, 17386 (1993)


\bibitem{read94}
N. Read, Sci. Tech. {\bf 9}, 1859 (1994); Surf. Sci. {\bf 361}, 7 (1996).


\bibitem{shankar97}
R. Shankar and G. Murthy, Phys. Rev. Lett. {\bf 79}, 4437 (1997).


\bibitem{lee98}
D.-H. Lee, Phys. Rev. Lett. {\bf 80}, 4547 (1998).


\bibitem{pasquier98}
V. Pasquier and F. D. M. Haldane, Nucl. Phys. {\bf B 516}, 719 (1998).


\bibitem{stern98}
B. I. Halperin and A. Stern, Phys. Rev. Lett. {\bf 80}, 5457 (1998);
A. Stern, B. I. Halperin, F. von Oppen, and S. Simon, 
Phys. Rev. B {\bf 59}, 12547 (1999).




\bibitem{read98} N. Read, Phys. Rev. B {\bf 58}, 16262 (1998).


\bibitem{shankar99} 
R. Shankar, Phys. Rev. Lett. {\bf 83}, 2382 (1999).


\bibitem{lilly98}
M. P. Lilly, J. P. Eisenstein, L. N. Pfeiffer, and K. W. West,
Phys. Rev. Lett. {\bf 80}, 1714 (1998).


\bibitem{ussishkin97}
I. Ussishkin and A. Stern, Phys. Rev. B {\bf 56}, 4013 (1997).


\bibitem{sakhi97}
S. Sakhi, Phys. Rev. B {\bf 56}, 4098 (1997).


\bibitem{kim96}
Y. B. Kim and A. J. Millis, Physica E {\bf 4}, 171 (1999);
preprint archived at cond-mat/9611125.


\bibitem{ussishkin98}
I. Ussishkin and A. Stern, Phys. Rev. Lett. {\bf 81}, 3932 (1998).


\bibitem{zhou99}
F. Zhou and Y. B. Kim, Phys. Rev. B {\bf 59} 7825 (1999).


\bibitem{zelakiewicz}
S. Zelakiewicz, H. Noh, T. J. Gramila, L. N. Pfeiffer, and K. W. West,
preprint archived at cond-mat/9907396.


\bibitem{bonsager98}
M. C. B{\o}nsager, K. Flensberg, B. Y.-K. Hu, and A. H. MacDonald,
Phys. Rev. B {\bf 57}, 7085 (1998); Physica B {\bf 249-251}, 864 (1998).


\bibitem{tso92} H. C. Tso, P. Vasilopoulos, and F. M. Peeters,
Phys. Rev. Lett. {\bf 68}, 2516 (1992).


\bibitem{zhang93} C. Zhang and Y. Takahashi, J. Phys.: Condens. Matter
{\bf 5}, 5009 (1993).


\bibitem{badalyan99} S. M. Badalyan and U. R\"{o}sler,
Phys. Rev. B {\bf 59}, 5643 (1999).


\bibitem{solomon89} P. M. Solomon, P. J. Price, D. J. Frank, and
D. C. La Tulipe, Phys. Rev. Lett. {\bf 63}, 2508 (1989).


\bibitem{gramila91}
T. J. Gramila, J. P. Eisenstein, A. H. MacDonald, L. N. Pfeiffer, and
K. W. West, Phys. Rev. Lett. {\bf 66}, 1216 (1991);
Surf. Sci. {\bf 263}, 446 (1992).


\bibitem{gramila93}
T. J. Gramila, J. P. Eisenstein, A. H. MacDonald, L. N. Pfeiffer, and
K. W. West, Phys. Rev. B {\bf 47}, 12957 (1993);
Physica B {\bf 197}, 442 (1994).


\bibitem{sivan92} U. Sivan, P. M. Solomon, and H. Shtrikman,
Phys. Rev. Lett. {\bf 68}, 1196 (1992).


\bibitem{jorger} C. J\"{o}rger, S. J. Cheng, H. Rubel, W. Dietsche,
R. Gerhardts, P. Specht, K. Eberl, and K. v. Klitzing,
preprint archived at cond-mat/9904214.


\bibitem{jauho93} A.-P. Jauho and H. Smith, Phys. Rev. B {\bf 47},
4420 (1993).


\bibitem{zheng93} L. Zheng and A. H. MacDonald, Phys. Rev. B {\bf 48},
8203 (1993).


\bibitem{kamenev95} A. Kamenev and Y. Oreg, Phys. Rev. B {\bf 52},
7516 (1995).


\bibitem{flensberg95} K. Flensberg, B. Y.-K. Hu, A.-P. Jauho, and J. Kinaret,
Phys. Rev. B {\bf 52}, 14761 (1995).


\bibitem{hu98} B. Y.-K. Hu, Phys. Rev. B {\bf 57}, 12345 (1998).


\bibitem{guven97} K. G\"{u}ven and B. Tanatar, Solid State Comm.
{\bf 104}, 439 (1997); Phys. Rev. B {\bf 56}, 7535 (1997);


\bibitem{price81} P. J. Price, Ann. Phys. (N.Y.) {\bf 133}, 217 (1981).


\bibitem{lyo88} S. K. Lyo, Phys. Rev. B {\bf 38}, 6345 (1988).


\bibitem{rojo99}
A. G. Rojo, J. Phys.: Cond. Mat. {\bf 11}, R31 (1999).


\bibitem{flensberg94}
K. Flensberg and B. Y.-K. Hu, Phys. Rev. Lett. {\bf 73}, 3572 (1994);
Phys. Rev. B {\bf 52}, 14796 (1995).


\bibitem{hill97}
N. P. R. Hill {\it et al.}, Phys. Rev. Lett. {\bf 78}, 2204 (1997).


\bibitem{noh98}
N. Noh, S. Zelakiewicz, X.-G. Feng, T. J. Gramila, L. N. Pfeiffer, and
K. W. West, Phys. Rev. B {\bf 58}, 12621 (1998).


\bibitem{khveshchenko}
D. V. Khveshchenko, preprint archived at cond-mat/9812093.


\bibitem{noh99}
N. Noh, S. Zelakiewicz, T. J. Gramila, L. N. Pfeiffer, and
K. W. West, Phys. Rev. B {\bf 59}, 13114 (1999).


\bibitem{melinte} S. Melinte et al., Phys. Rev. Lett. {\bf 84}, 354 (2000).


\bibitem{dementyev} A. E. Dementyev et al., Phys. Rev. Lett. {\bf 83}, 
5074 (1999).


\bibitem{kim94} Y. B. Kim, A. Furusaki, X.-G. Wen, and P. A. Lee, 
Phys. Rev. B {\bf 50}, 17917 (1994).


\bibitem{ioffe94} B. L. Altshuler, L. B. Ioffe, and A. J. Millis, 
Phys. Rev. B {\bf 50}, 14048 (1994).


\bibitem{bkm} M. C. B{\o}nsager, Y. B. Kim, and A. H. MacDonald, work in progress. 


\bibitem{stern67} F. Stern, Phys. Rev. Lett. {\bf 18}, 546 (1967).


\bibitem{fermidirac}
We have used a FORTRAN code authored by Allan MacLeod and Michele Goano
available from http://www.netlib.org
See also, M. Goano, Solid-State Electronics, {\bf 36}, 217 (1993).  


\end{thebibliography}
\end{document}